\documentclass[12pt]{article}
\usepackage{amssymb,amsmath,epsfig}

\begin{document}

\title{\bf Born-Infeld Thin-shell Wormholes Supported by Generalized Cosmic Chaplygin Gas}

\author{M. Azam \thanks{azam.math@ue.edu.pk}\\
Division of Science and Technology, University of Education,\\
Township Campus, Lahore-54590, Pakistan.}

\date{}

\maketitle
\begin{abstract}
This paper investigates thin-shell wormholes in Born-Infeld theory
supported by generalized Cosmic Chaplygin gas (GCCG). We study their
stability via radial perturbations for distinct values of charge and
Born-Infeld parameter. The comparison of wormhole solutions
corresponding to generalized Chaplygin gas, modified Chaplygin gas
with GCCG quation of state is established. It is found that similar
type of wormhole solutions exists for small value of charge and
Born-Infeld parameter for all type of equation of state, while some
extra stable as well as unstable solution are found corresponding to
large value of charge and Born-Infeld parameter. Thus, it is
concluded that GCCG and large value of charge may responsible for
such extra solutions.
\end{abstract}

\section{Introduction}

A wormhole (WH) is a hypothetical object in spacetime which behaves
like a smooth bridge between two different universes or smooth
shortcut between remote parts of a single universe (Visser, 1989).
Recently, WH physics has been taken as hot cake due to its
interesting features. The first traversable WH was obtained by
Morris and Thorn (1988) with two asymptotically flat regions which
are joined by a minimal surface area known as WH throat. For WH to
be traversable, it is necessary that the WH throat must satisfies
the flare out condition (Lobo, 2008), i.e., matter threading the WH
throat should violate the null energy condition. Such matter is
known as exotic matter. However, physical viability of such WHs was
a big issue. In this scenario, Visser (2003) showed that the amount
of exotic matter located around the throat can be reduced by taking
an appropriate choice of WH geometry.

It is well known that the existence of exotic matter (violation of
null energy condition) is always accompanied by WH solutions. In
this context, one can construct thin-shell WH through cut and paste
technique to confine the exotic matter at WH throat (Lobo, 2008).
This is an elegant and efficient procedure used to minimize the
violation of null energy condition through thin-shell formalism.
Poisson and Visser (1995) were the pioneer who constructed
Schwarzschild thin-shell WH and investigated its stability through
radial perturbations. After that, many authors (Lobo, 2005; Lemos
and Lobo, 2008; Rahaman et al. 2009; Goncalo, 2010; Rahaman et al.
2012) have constructed thin-shell WHs with this procedure and used
Darmois-Israel junction conditions (Darmois, 1927; Israel, 1966,
ibid. 1967) to explore WH dynamics. In recent years, some authors
(Eiroa and Simeone, 2005; Eiroa, 2008; Rahaman et al. 2010; Bejarano
and Eiroa, 2011) constructed spherical thin-shell WHs in this theory
and investigated their linearized stability under radial
perturbations.

The selection of equation of state (EoS) for dynamical analysis of
matter present at the shell has an important role in the existence
and stability of WH solutions. Thus, many authors have taken account
different EoS in search of viable thin-shell WHs. In this context,
family of Chaplygin gas (Chaplygin (1939); Von Karman (1941)) has
been used successfully in describing various astronomical phenomenon
(wormholes, cosmological evolution of the early and present
Universe). Eiroa and Simeone (2007) found stable static WH solutions
with Chaplygin gas corresponding to fixed values of parameters.
Later on, Bandyopadhyay et al. (2009) carried out this work with
simple modified Chaplygin gas (MCG) EoS and found some more stable
WH solutions. Eiroa (2009) have constructed thin-shell WHs
numerically supported with generalize Chaplygin gas (GCG) and shows
that some extra solution can exists. This indicates that choice of
EoS may play a significant role in the existence of WH solutions.
Also, Gorini et al. (2008, 2009) found WH like solutions by using
Chaplygin gas and GCG. We have also studied thin-shell WHs with
family of Chaplygin gas and found distinct solutions corresponding
to distinct EoS (Sharif and Azam, 2013a, 2013b, 2013c).

Since GCCG is less constrained as compared to MCG and GCG and is
capable of adapting itself to any domain of cosmology, depending
upon the choice of parameters. Thus it has a more universal
character and the big-rip singularity can easily be avoided in this
model. For instance, many authors have used GCCG to; discuss the
evolution of the universe from dust era to $\Lambda$CDM (Chakraborty
et al. 2007), studied the background dynamics of GCCG in brane world
gravity (Rudra, 2012), presented a singularity free model for an
expanding universe undergoing a late acceleration (Ratul et al.
2013), studied the role of GCCG in accelerating universe (Prabir,
2013), discussed FRW universe in loop Quantum gravity with GCCG as
dark energy candidate (Ranjit and Debnath, 2014), studied GCCG
inflationary universe model for a flat FRW geometry (Sharif and
Rabia, 2014), found stable traversable WHs (Sharif and Jawad, 2014),
found stable and unstable Schwarzschild de-sitter and anti-de-sitter
thin-shell WHs (Sharif and Mumtaz, 2014).

The nonlinear electrodynamics (NED) theory which is considered as
the most outstanding viable theory among all the NED theories
introduced by Born-Infeld (BI) in $1934$ (Born and Infeld, 1934). It
has a therapeutic power for singularities appear in Maxwell theory.
Hoffmann (1935) found spherically symmetric solution by coupling
general relativity with BI electrodynamics theory which describes
the gravitational field of a charged object. It was shown that
Maxwell and BI theories possess the property of duality invariance
like electric and magnetic fields (Gibbons and Rasheed, 1995). The
use of BI action in the low energy string theory has been
interesting to study such NED theories (Fradkin and Tseytlin, 1985;
Bergshoeff et al. 1987; Metsaev et al. 1987; Tseytlin, 1997; Brecher
et al. 1998). It was argued that trajectories of photons are not
null geodesics of the background metric in curved spacetimes within
BI electrodynamics (Pleba\~{n}ski, 1970) but rather follows null
geodesics of a physical geometry influenced by the nonlinearities of
electromagnetic field. Bret´on (2002) examined the geodesic
structure of BI black holes.

The Born-Infeld action in four-dimension associated with Einstein
gravity is given by
\begin{equation}\label{1a}
\mathcal{S}=\int{d^4x}\sqrt{g}\left(\frac{R}{16\pi}+L_{BI}\right),
\end{equation}
where $g,~R$ and $L$ correspond to the determinant of the metric
tensor, Ricci scalar and non-linear Lagrangian coupled with
electromagnetic field tensor defined as
\begin{equation}\label{2a}
L_{BI}=\frac{1}{4\pi{b^2}}\left(1-\sqrt{1+\frac{1}{2}F_{\sigma\nu}F^{\sigma\nu}b^2-
\frac{1}{4}{^\ast}F_{\sigma\nu}F^{\sigma\nu}b^4}\right),
\end{equation}
where
$F_{\sigma\nu}=\partial_{\sigma}A_{\nu}-\partial_{\nu}A_{\sigma}$
and
$^{\ast}{F_{\sigma\nu}}=\frac{1}{2}\sqrt{-g}\epsilon_{\gamma\delta\sigma\nu}F^{\gamma\delta}$
are electromagnetic field tensor and Hodge dual of $F_{\sigma\nu}$,
respectively and $\epsilon_{\gamma\delta\sigma\nu}$ is the
Levi-Civita symbol. The value of BI parameter $b$ will make a
comparison between Born-Infeld and Maxwell electrodynamics, i.e., BI
Lagrangian will approach to Maxwell Lagrangian in the limit
$b\rightarrow{0}$. The variation of action with respect to
$g_{\mu\nu}$ and $A_\nu$ yields Einstein field equations whose
solution corresponds to vacuum spherically symmetric solution
(Gibbons and Rasheed, 1995; Bret´on, 2002) given by
\begin{equation}\label{1}
ds^2=-H(r)dt^{2}+H^{-1}(r)dr^{2}+r^2(d\theta^{2}
+\sin^2\theta{d\phi^2}),
\end{equation}
with
\begin{equation}\label{2}
H(r)=1-\frac{2M}{r}+\frac{2}{3b^2}\left\{r^2-\sqrt{r^4+b^2Q^2}+
\frac{\sqrt{|bQ|^3}}{r}F\left[{\arccos}\left(\frac{r^2-|bQ|}{r^2+|bQ|}\right),\frac{\sqrt{2}}{2}\right]\right\},
\end{equation}
where elliptic integral of the first kind $F(\gamma,k)$ is defined
by
$$F(\gamma,k)=\int^{\text{sin}{\gamma}}_0[(1-y^2)(1-k^2y^2)]^{\frac{-1}{2}}dy=
\int^{\text{sin}{\gamma}}_0(1-k^2sin^2{\phi})^{\frac{-1}{2}}d\phi,$$
and $M,~Q$ represent mass and charge of the BI black hole. The
horizons of (\ref{1}) can be found numerically by setting $H(r)=0$.
A regular event horizon is obtained for a given value of $b$ and
small value of charge, i.e., $0\leq{\frac{|Q|}{M}}\leq{\omega_1}$,
where
$\omega_1=(\frac{9|b|}{M})^\frac{1}{3}[F(\pi,\frac{\sqrt{2}}{2})]^\frac{-2}{3}$.
For $\omega_1<{\frac{|Q|}{M}}<{\omega_2}$, there exist two regular
horizons inner and outer similar to Reissner-Nordstr\"{o}m geometry.
However, for $\frac{|Q|}{M}=\omega_2$ and $\frac{|Q|}{M}>\omega_2$,
there exists one degenerate horizon and naked singularity
respectively, where $\omega_2$ can be obtained numerically through
the condition $H(r)=0=H'(r)$ (see details (Eiroa and Aguirre, 2012).
It is easy to check that in the limit $b\rightarrow{0}$
(Reissner-Nordstr\"{o}m case), $\omega_1=0$ and $\omega_2=1$.

Recently, much interest in non-linear electrodynamics theories has
been aroused in application to WH geometries and cosmological
phenomena (Mazharimousavi and Halilsoy, 2015; Gullu et al. 2015;
Jana and Kar, 2015; Gullu et al. 2015; Mazharimousavi et al. 2013).
In this scenario, Baldovin et al. (2000) showed that a certain field
configuration in Born-Infeld electromagnetism in flat spacetime can
be interpreted as a WH. Arellano et al. (2009) studied some
properties for the evolving WH solutions in non-linear
electrodynamics. Richarte and simeone (2009, 2010) studied the
spherically symmetric thin-shell WHs in the scenario of Born-Infeld
gravity and analyzed the mechanical stability of WH configurations.
Rahaman et al. (2010) construct and discuss various aspects of
thin-shell WHs from a regular charged black hole in the framework of
non-linear electrodynamics. Eiroa and Simeone (2011) investigated
the mechanically stability of thin-shells both in Einstein Maxwell
and Born Infeld theory. Mazharimousavi et al. (2011) used the
Hoffman-Born-Infeld Lagrangian to construct the black holes and
viable thin-shell WHs. In particular, they investigate the stability
of thin-shell WHs supported by normal matter. Halilsoy et al. (2014)
constructed thin-shell WHs from the regular Hayward black hole with
linear, logarithmic, Chaplygin etc., equation of states and found
that Hayward parameter makes thin-shell WHs more stable.

In this work, we have construct BI thin-shell WHs and investigate
their stability supported with GCCG. We have compared our results
with recent work supported with GCG (Eiroa and Aguirre, 2012) and
MCG (Sharif and Azam, 2014). The paper is planned as: Section
\textbf{2} deals with the basic equations for the construction of
spherical thin-shell WHs with GCCG. In section \textbf{3}, we
present the general procedure to investigate stability of static WH
solutions. In section \textbf{4}, we apply the general formalism
developed in section \textbf{2} and \textbf{3} to construct and
explore stability of BI thin-shell WHs. We summarize our results in
the last section.

\section{Thin-Shell Wormhole Construction: Fundamental Equations}

In this section, we will work out basic equations for BI thin-shell
WH with GCCG. We take two copies
$,\mathcal{S}^{\pm}=\{x^{\mu}=(t,r,\theta,\phi)/r\geq{a}\},$ of the
BI black hole (\ref{1}) each with $r\geq{a}$ in such a way that
these geometries are prevented from horizons and singularities,
where $`a`$ is a constant. The joining of these geometries at the
timelike hypersurfaces $\Sigma=\Sigma^\pm=\{x^\mu/r-a=0\}$, to form
a new manifold (geodesically complete)
$\mathcal{S}^{\pm}=\mathcal{S}^{+}\cup\mathcal{S}^{-}$ representing
a WH having two regions connected by a junction surface (throat of
the WH).

For the dynamical analysis of this traversable WH, we shall adopt
the Darmois-Israel formalism. This formalism is one of the basic
formulation used to study the dynamics of the matter field located
at the WH throat, providing a set of equations correspond to field
equations. The synchronous timelike hypersurface which is throat of
the WH (junction surface) is defined by the coordinates
$\varsigma^i=(\tau,\theta,\phi)$. Thus the intrinsic metric to the
hpersurface $\Sigma$ can be written as
\begin{equation}\label{4}
ds^2=-d\tau^2+a^2(\tau)(d\theta^2+\sin^2\theta{d\phi^2}),
\end{equation}
where $\tau$ is the proper time. Following the Darmois-Israel
formalism, the explicit expression for the extrinsic curvature
(second fundamental forms) connected with two sides of the shell is
defined as
\begin{equation}\label{5}
K^{\pm}_{ij}=-n^{\pm}_{\gamma}\left(\frac{{\partial}^2x^{\gamma}_{\pm}}
{{\partial}{\varsigma}^i{\partial}{\varsigma}^j}+{\Gamma}^{\gamma}_{{\mu}{\nu}}
\frac{{{\partial}x^{\mu}_{\pm}}{{\partial}x^{\nu}_{\pm}}}
{{\partial}{\varsigma}^i{\partial}{\varsigma}^j}\right),\quad(i,~j=\tau,\theta,\phi),
\end{equation}
where the superscripts $\pm$ stands for exterior and interior
geometry, respectively. The outwards unit 4-normals to $\Sigma$ with
,$n^{\gamma}n_{\gamma}=+1$, are given by
\begin{equation}\label{6}
n^{\pm}_{\gamma}=\pm\left|g^{\mu\nu}\frac{\partial{f}}{\partial{x^{\mu}}}
\frac{\partial{f}}{\partial{x^{\nu}}}\right|^{-\frac{1}{2}}\frac{\partial{f}}{\partial{x^\gamma}}
=\left(-\dot{a},\frac{\sqrt{H(r)+\dot{a}^2}}{H(r)},0,0\right).
\end{equation}
Using Eqs. (\ref{4}) and (\ref{5}), we obtain the non-vanishing
components for geometry (\ref{1}) as follows
\begin{equation}\label{7}
K^{\pm}_{\tau\tau}=\mp\frac{H'(a)+2\ddot{a}}{2\sqrt{H(a)+\dot{a}^2}},
\quad
K^{\pm}_{\theta\theta}=K^{\pm}_{\phi\phi}={\pm}\frac{1}{a}\sqrt{H(a)+\dot{a}^2},
\end{equation}
where prime and dot corresponds to $\frac{d}{dr}$ and
$\frac{d}{d\tau}$, respectively.

The surface stresses, i.e., surface energy density $\sigma$ and
surface pressures $p=p_\theta=p_\phi$, are determined by the surface
stress-energy tensor
$S_{ij}=\text{diag}(\sigma,p_{\theta},p_{\phi})$ and Einstein
equations or Lanczos equations on the shell are given by
\begin{equation}\label{8}
S_{ij}=\frac{1}{8\pi}\left\{g_{ij}K-[K_{ij}]\right\},
\end{equation}
where $$[K_{ij}]=K^{+}_{ij}-K^{-}_{ij},\quad
K=tr[K_{ij}]=[K^{i}_{i}].$$ From Eqs. (\ref{7}) and (\ref{8}), the
surface stresses of the shell turns out as
\begin{eqnarray}\label{9}
\sigma&=&-\frac{\sqrt{H(a)+\dot{a}^2}}{2\pi{a}},\\\label{10}
p&=&p_{\theta}=p_{\phi}=\frac{\sqrt{H(a)+\dot{a}^2}}{8\pi}
\left[\frac{2\ddot{a}+H'(a)}{H(a)+\dot{a}^2}+\frac{2}{a}\right].
\end{eqnarray}
From the above equation, the negativity of surface energy density
will insure the presence of exotic matter at the throat. To discuss
the physical aspects of this exotic matter, we choose GCCG
(Gonz\'{a}lez-Diaz, 2003) EoS because the free parameter in it can
encompass different types of matter defined by
\begin{eqnarray}\label{10a}
p=-\frac{1}{\sigma^{\beta}}\left[L+(\sigma^{1+\beta}-L)^{-\omega}\right],
\end{eqnarray}
where $L=\frac{D}{1+\omega}-1$, $D\in(-\infty,\infty)$ and
$-D<\omega<0$. Here, we take $D$ to be positive constant other than
unity. Also, the above equation reduces to GCG in the limit
$\omega\rightarrow{0}$.

For the dynamical analysis of thin-shell WH, we develop a second
order differential equation (equation of motion) by using
Eqs.(\ref{9}) and (\ref{10}) in (\ref{10a}) given by
\begin{eqnarray}\nonumber
&&\left\{\left[2\ddot{a}+H'(a)\right]a^2+\left[H(a)+\dot{a}^2
\right]2a\right\}\left[2a\right]^{\beta}-2
(4\pi{a^2})^{1+\beta}\left[H(a)+\dot{a}^2\right]^\frac{1-\beta}{2}\\\label{11}
&\times&\left[L+\left\{(2\pi{a})^{-(1+\beta)}
(H(a)+\dot{a}^2)^{\frac{(1+\beta)}{2}}-L\right\}^{-\omega}\right]
=0.
\end{eqnarray}
This equation provides full understanding of the thin-shell WH
satisfied by the throat radius with GCCG in BI theory.

\section{Stability Analysis of Static Solutions: A Standard Approach}

In this section, we follow the standard approach to investigate the
stability of static BI WH solutions with GCCG under radial
perturbations (Bejarano and Eiroa, 2011). From Eqs.(\ref{9}),
(\ref{10}) and (\ref{11}), the static configuration of surface
energy density, surface pressure and evolution equation of BI WH
takes the form
\begin{eqnarray}\label{13}
\sigma_0=-\frac{\sqrt{H(a_0)}}{2\pi{a_0}},\quad
p_0=\frac{2H(a_0)+a_0H'(a_0)}{8\pi{a_0}\sqrt{H(a_0)}},
\end{eqnarray}
\begin{eqnarray}\nonumber
&&\left[a^2_0H'(a_0)+2a_0H(a_0)
\right]\left[2a_0\right]^{\beta}-2(4\pi{a_0^2})^{1+\beta}\left[H(a_0)\right]^\frac{1-\beta}{2}\\\label{14}&\times&
\left[L+\left\{(2\pi{a_0})^{-(1+\beta)}
\left(H(a_0)\right)^{\frac{(1+\beta)}{2}}-L\right\}^{-\omega}\right]=0.
\end{eqnarray}
The law of conservation of energy on the WH throat can be defined
with Eqs.(\ref{9}) and (\ref{10}) as
\begin{eqnarray}\label{15}
\frac{d}{d\tau}(\sigma{\Omega})+p\frac{d\Omega}{d\tau}=0,
\end{eqnarray}
where $\Omega=4\pi{a^2}$ is known as area of WH throat. This
equation describes that the sum of rate of change of WH throat's
internal energy and work done by the internal forces is equal to
zero. The above equation can be written as
\begin{eqnarray}\label{16}
\dot{\sigma}=-2(\sigma+p)\frac{\dot{a}}{a},
\end{eqnarray}
and using ${\sigma}'=\frac{\dot{\sigma}}{\dot{a}}$, it yields
\begin{equation}\label{17}
a{\sigma}'=-2(\sigma+p).
\end{equation}
The integral solution of Eq.(\ref{16}) provides the full
understanding of WH dynamics given as
\begin{eqnarray}\label{17a}
\ln\frac{a}{a(\tau_0)}=-\frac{1}{2}\int^\sigma_{\sigma(\tau_0)}{\frac{d\sigma}{\sigma+p(\sigma)}},
\end{eqnarray}
and can then be inverted to obtain $\sigma=\sigma(a)$. Thus,
Eq.(\ref{9}) takes the form
\begin{equation}\label{18}
\dot{a}^2+\phi(a)=0,
\end{equation}
where $\phi(a)$ is the potential function
\begin{equation}\label{19}
\phi(a)=H(a)-\left[2\pi{a}{\sigma(a)}\right]^2.
\end{equation}
Taking the derivative of above equation and using (\ref{17}), we
have
\begin{equation}\label{19a}
\phi'(a)=H'(a)+8{\pi}^2a\sigma(a)\left[\sigma(a)+p(a)\right].
\end{equation}
This potential function is used to discuss the linearized stability
analysis of static solution under radial perturbations. For this
purpose, we apply Taylor expansion to potential function upto second
order around $a=a_0$ to provides
\begin{eqnarray}\label{20}
\phi(a)=\phi(a_0)+\phi'(a_0)(a-a_0)+\frac{1}{2}\phi''(a_0)(a-a_0)^2+O[(a-a_0)^3].
\end{eqnarray}
The first derivative of EoS takes the form
\begin{equation}\label{21}
p'(a)=\sigma'(a)\left[\omega(1+\beta)(\sigma(a)^{1+\beta}-L)^{-1-\omega}-\frac{\beta{p}}{\sigma}\right],
\end{equation}
which further can be written as
\begin{equation}\label{22}
\sigma'(a)+2p'(a)=\sigma'(a)\left[1+2\omega(1+\beta)(\sigma(a)^{1+\beta}-L)^{-1-\omega}
-\frac{2\beta{p(a)}}{\sigma(a)}\right].
\end{equation}
The stability of static WH solutions needs conditions
$\phi(a_0)=0=\phi'(a_0)$, also stable and unstable WH solutions
corresponds to $\phi''(a_0)>0$ and $\phi''(a_0)<0$, respectively. It
can be easily verified that $\phi(a_0)=0=\phi'(a_0)$ by using
Eq.(\ref{13}) in Eqs.(\ref{19}) and (\ref{19a}). Now, the second
derivative of the potential function with Eq.(\ref{22}) can be
written as
\begin{eqnarray}\nonumber
\phi''(a)&=&H''(a)-8{\pi}^2\left\{[\sigma(a)+2p(a)]^2+2\sigma(a)
(\sigma(a)+p(a))\left[\left(1-2\beta\frac{p}{\sigma}\right)\right.\right.
\\\label{24}&+&\left.\left.+2\omega(1+\beta)(\sigma^{1+\beta}-L)^{-1-\omega}\right]\right\},
\end{eqnarray}
and using Eq.(\ref{13}) in the above equation, it leads to
\begin{eqnarray}\nonumber
\phi''(a_0)&=&H''(a_0)+\frac{(\beta-1)[{H'(a_0)}]^2}{2H(a_0)}+\frac{H'(a_0)}{a_0}
\left\{1-2\omega(1+\beta)\right.\\\nonumber&\times&\left.
\left[\left(\frac{\sqrt{H(a_0)}}{2\pi{a_0}}\right)^{1+\beta}+L\right]^{-1-\omega}\right\}
-\frac{2H(a_0)(1+\beta)}{a^2_0}\\\label{25}&\times&\left\{1-2\omega
\left[\left(\frac{\sqrt{H(a_0)}}{2\pi{a_0}}\right)^{1+\beta}+L\right]^{-1-\omega}\right\}.
\end{eqnarray}

\section{Born-Infeld Thin-Shell Wormholes}

This section deals with the possible existence of BI thin-shell WHs
with GCCG corresponding to various values of $M$, $Q$, BI parameter
$b$ of BI black hole and constant $L$. Each numerical solution of
Eq.(\ref{14}) for $a_0$ will represent a BI thin-shell WH. We
replace these solutions in Eq.(\ref{25}) to investigate its
stability. For this purpose, the whole region is divided into three
parts: if the contour lies in the regions where $a_0>r_h$ and
$\phi''(a_0)>0$ or $\phi''(a_0)<0$ will correspond to the stable
(light green) or unstable (light yellow) static solution and will be
non-physical zone (grey) if $a_0\leq{r_h}$, where $r_h$ is event
horizon of BI black hole.
\begin{figure}
\centering \epsfig{file=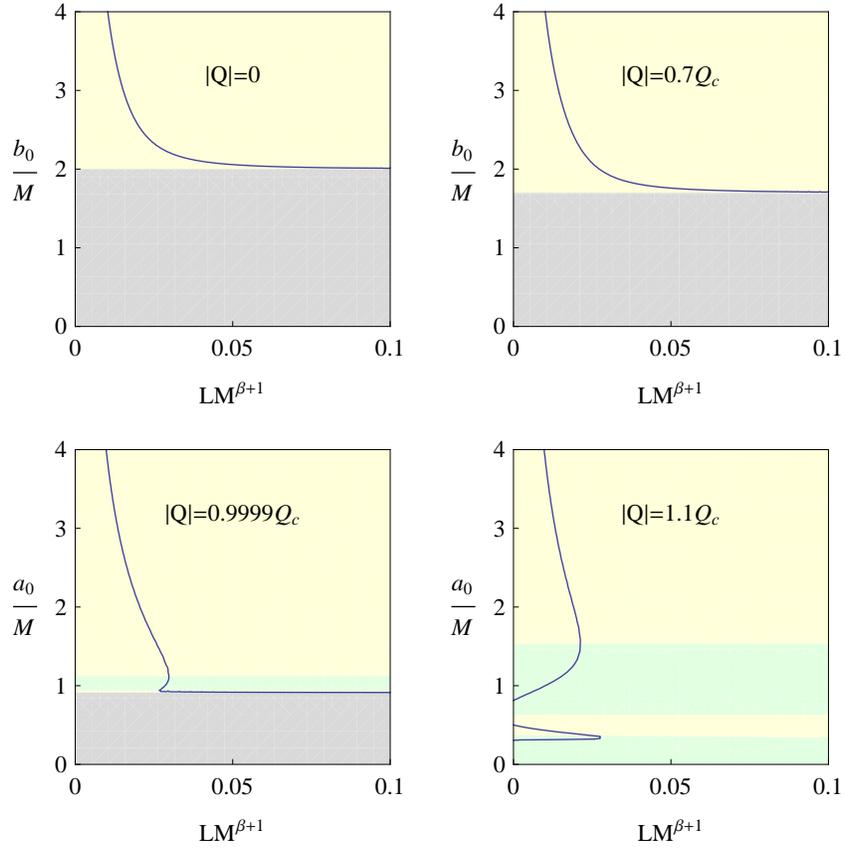}\caption{Thin-Shell WHs with
Born-Infeld parameter $\frac{b}{M}=1$, $L=0.09, \omega=-5$, gas
exponent $\beta=0.2$ and distinct values of charge.}
\end{figure}
It is noted that static solutions have an important change around
$\frac{Q_c}{M}$ depending upon the BI parameter $\frac{b}{M}$, where
$Q_c$ is the critical charge associated with $Q$ for which the given
metric has no horizon. We have chosen those values of
$\frac{Q_c}{M}$ for plotting for which $H(r_h)=0=H'(r_h)$ (obtained
numerically). We find that event horizon radius $\frac{a_0}{M}$
decreases as charge increases and finally disappear when
$\frac{Q}{M}>\frac{Q_c}{M}$ (Figures \textbf{1}-\textbf{6}). The
results of Eqs.(\ref{14}) and (\ref{25}) for different parametric
values of $\frac{b}{M}=1,~2,~5$, $\beta=0.2,~1$, charge and constant
$L$ are summarizes as follows.
\begin{figure}
\centering \epsfig{file=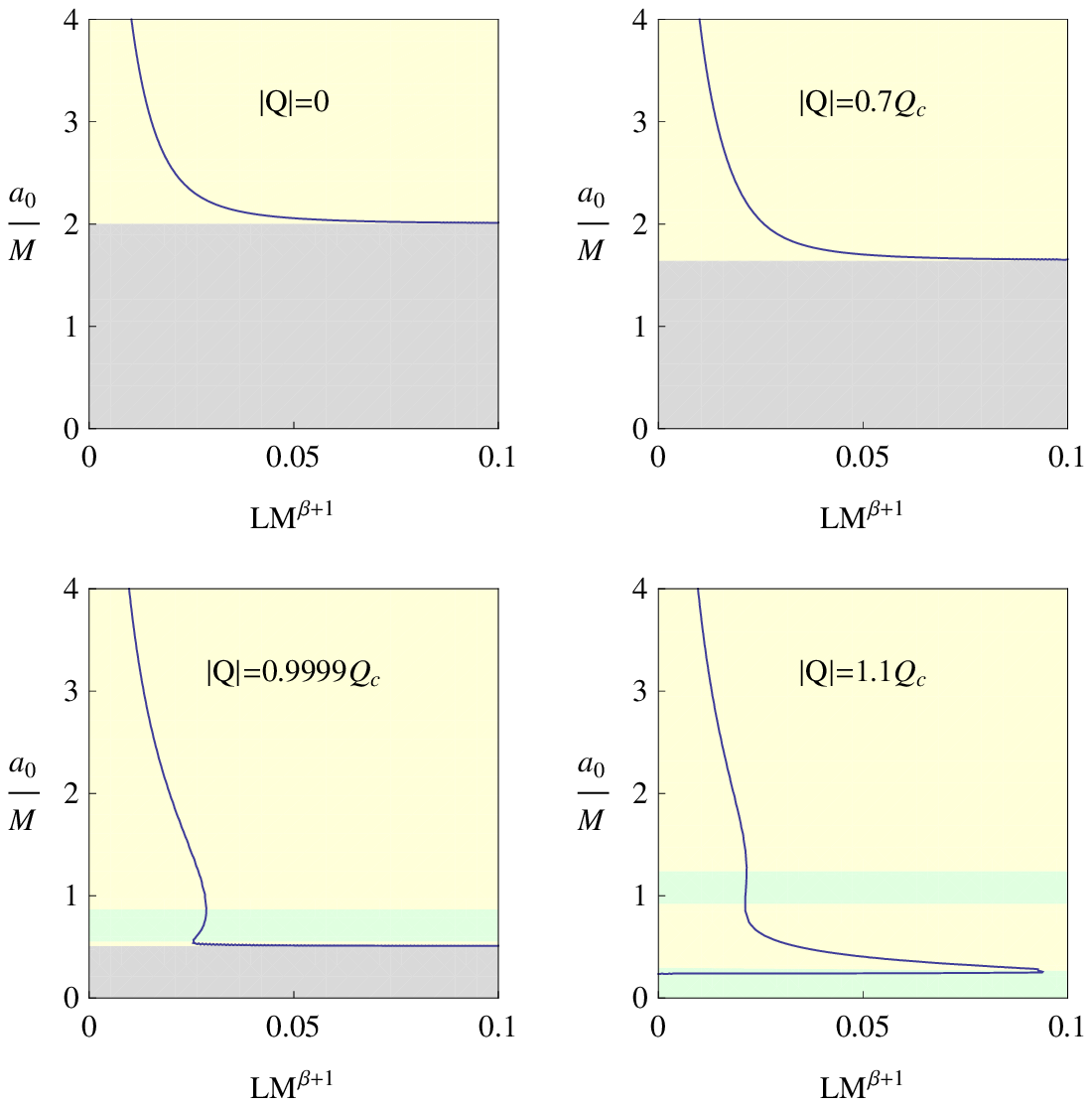}\caption{For $\frac{b}{M}=2$,
$\beta=0.2$, $L=0.09, \omega=-5$ and distinct values of charge.}
\end{figure}
\begin{figure}
\centering \epsfig{file=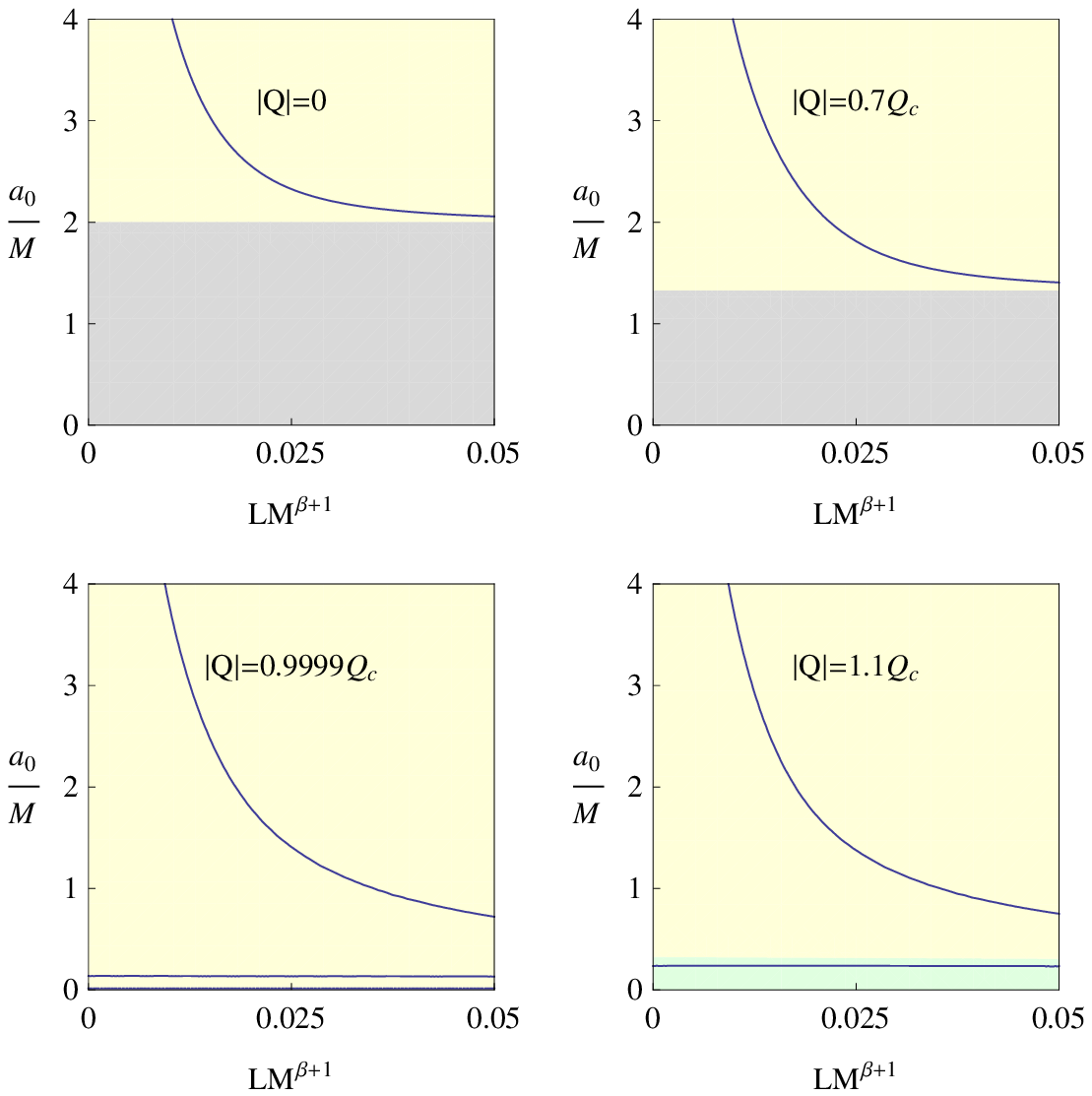}\caption{For $\frac{b}{M}=5$,
$\beta=0.2$, $L=0.09, \omega=-5$ and distinct values of charge.}
\end{figure}
\begin{figure}
\centering \epsfig{file=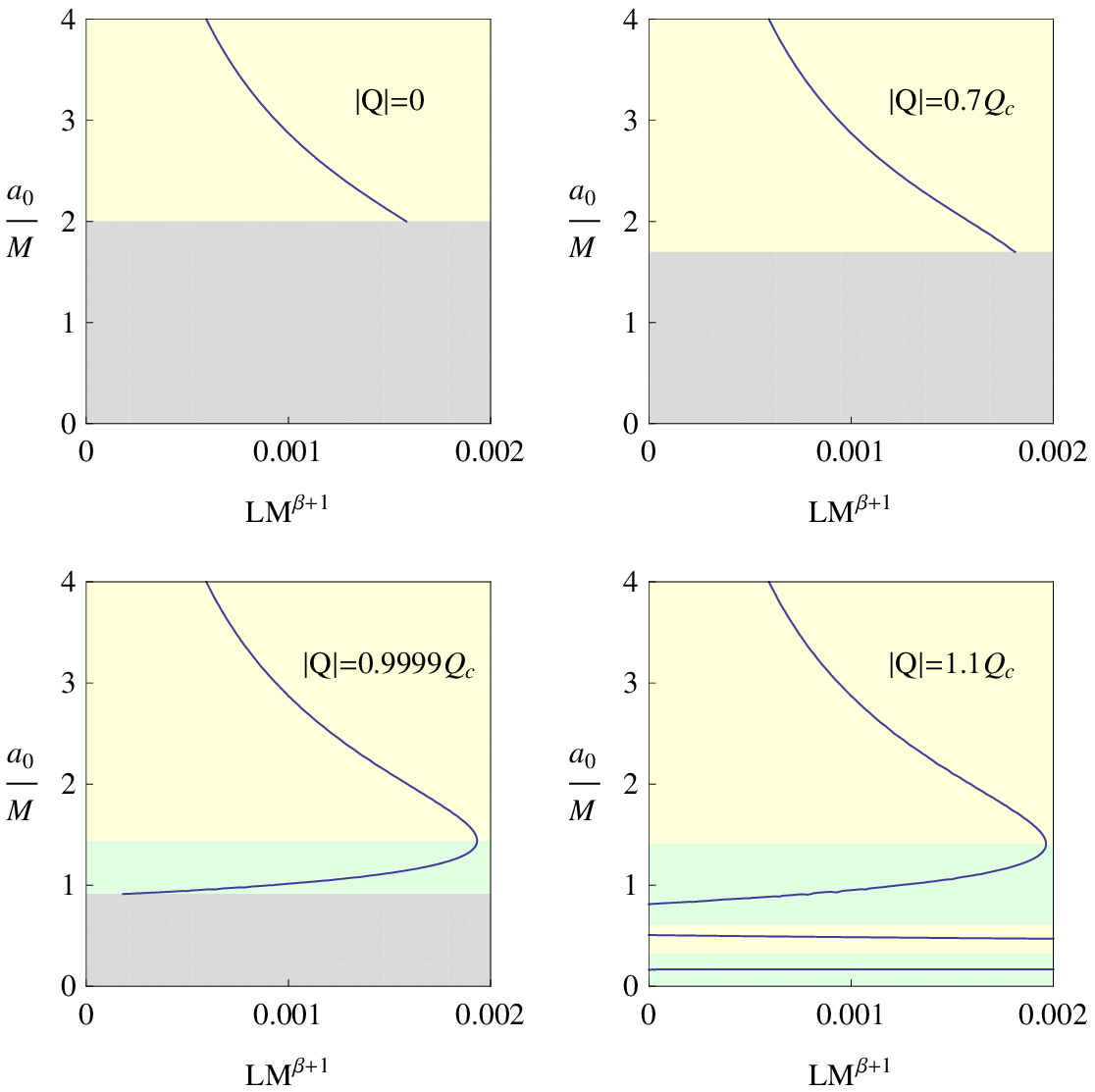}\caption{For $\frac{b}{M}=1$,
$\beta=1$, $L=0.09, \omega=-5$ and distinct values of charge.}
\end{figure}
\begin{figure}
\centering \epsfig{file=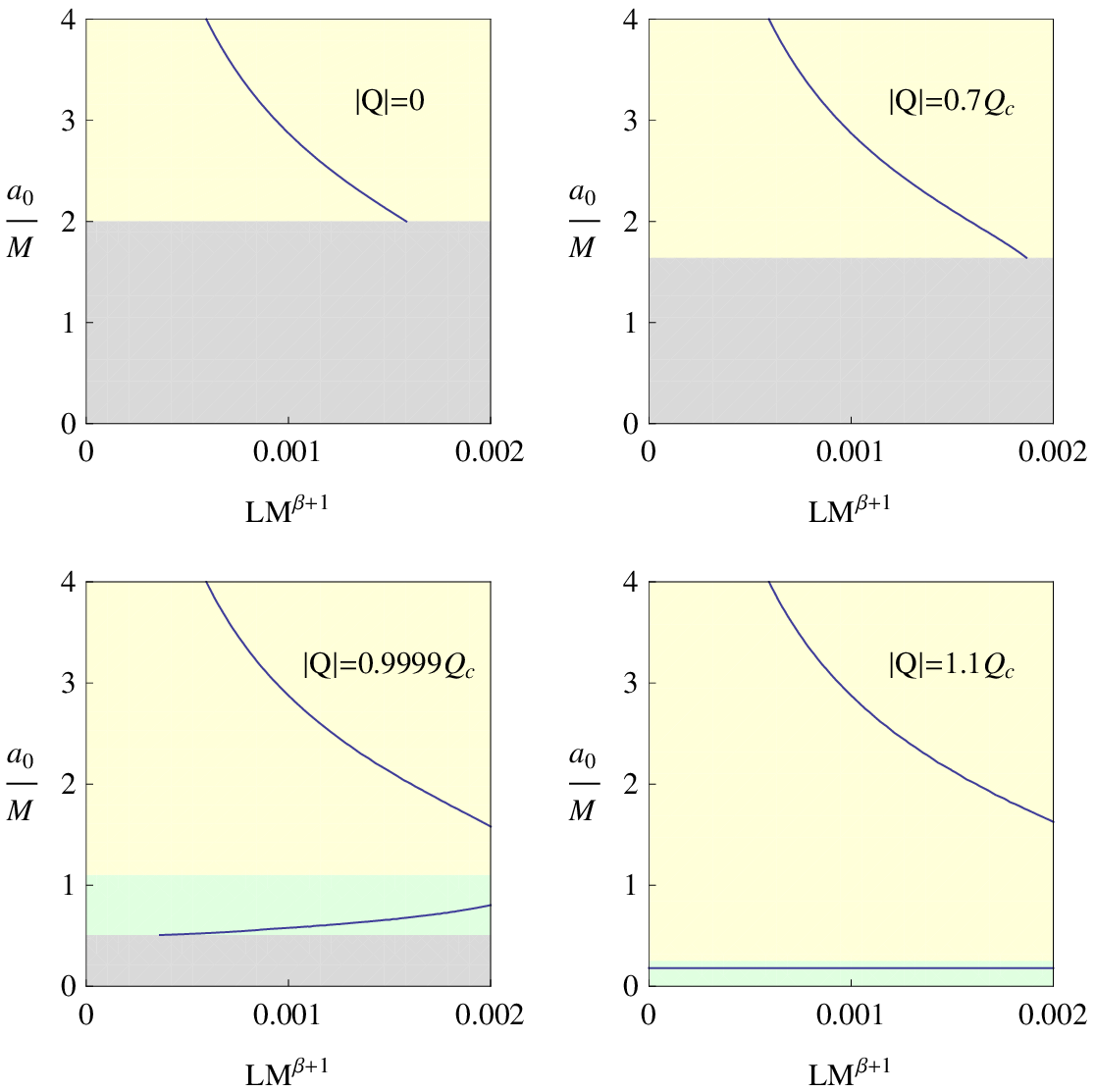}\caption{For $\frac{b}{M}=2$,
$L=0.09, \omega=-5$, $\beta=1$, and distinct values of charge.}
\end{figure}
\begin{figure}
\centering \epsfig{file=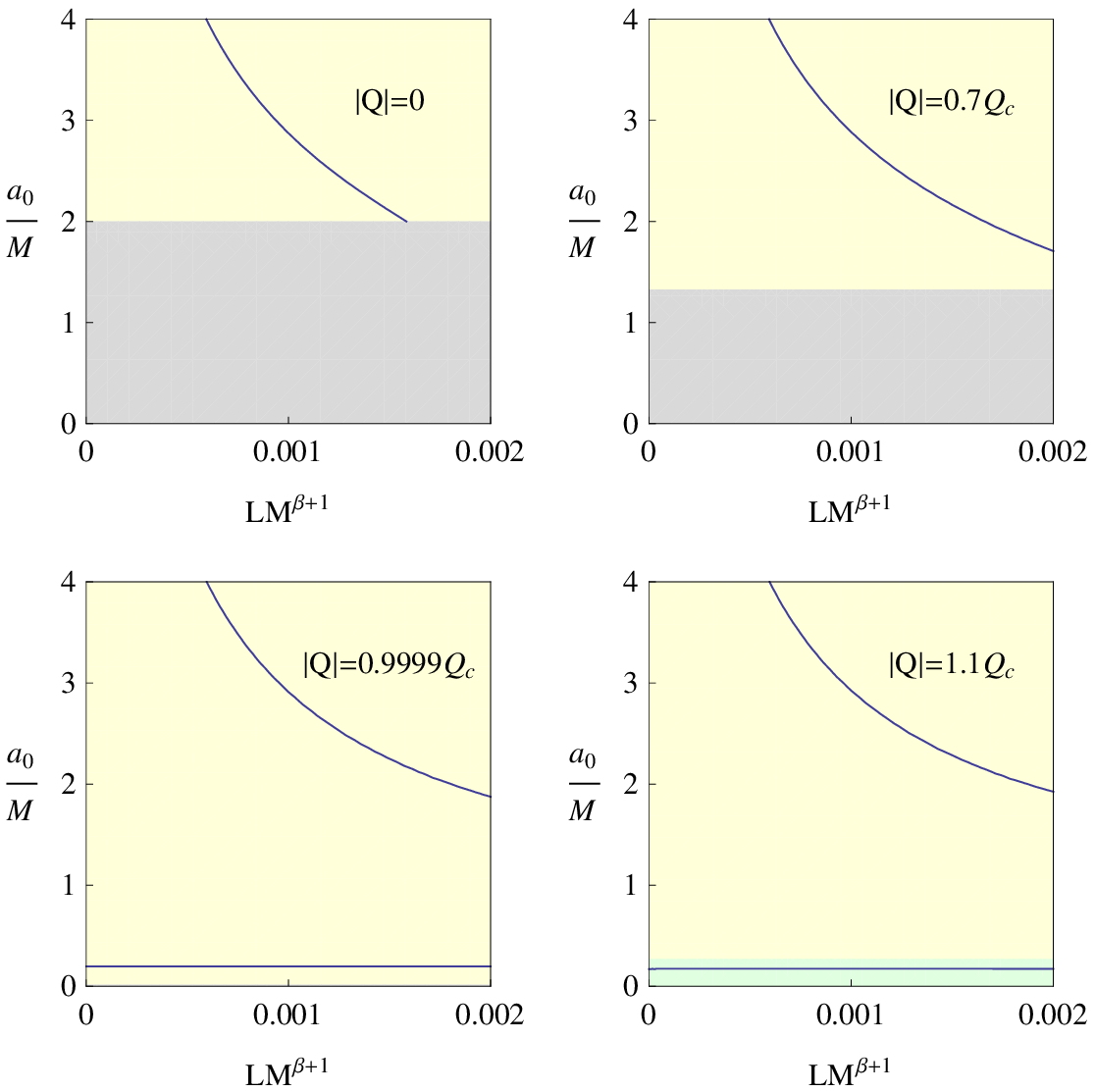}\caption{For $\frac{b}{M}=5$,
$\beta=1$, $L=0.09, \omega=-5$ and distinct values of charge.}
\end{figure}
\begin{itemize}
\item{Figure \textbf{1,~2,~3}, shows static WH solutions
corresponding to gas exponent $\beta=0.2$, different values of
charge and BI parameter $\frac{b}{M}=1,2,5$. We see that there exist
one unstable solution for each case corresponding to $|Q|=0$ and
$|Q|=0.7Q_c$ with $\frac{b}{M}=1,2,5$. When $\frac{b}{M}=1,2$, we
observe that there are two unstable and one stable solutions
corresponds to $|Q|=0.9999Q_c$ and two (unstable and stable)
solutions corresponds to $|Q|=1.1Q_c$, while for $\frac{b}{M}=5$
only stable and unstable solutions are exist corresponding to
$|Q|=0.9999Q_c$ and $|Q|=1.1Q_c$, respectively. In each case, the
critical charge has different value when $\beta=0.2$, i.e.,
$\frac{Q_c}{M}=1.02526,~1.10592,~1.148468$ for $\frac{b}{M}=1,2,5$,
respectively. Moreover, the horizon of the original manifold
decreases continuously for increasing value of charge and eventually
disappears for large value of $\frac{b}{M}$ and $|Q|>Q_c$, where
both unstable and stable solutions are exists.}
\item{Figures \textbf{4,~5,~6}, represents static WH solutions
with $\beta=1$, $\frac{b}{M}=1,2,5$ and different values of charge.
The possible solutions are similar to the above case for
$|Q|=0,~0.7Q_c$ and $\frac{b}{M}=1,2,5$, while both stable and
unstable solutions exist when $\frac{b}{M}=1,2$ and $|Q|=0.9999Q_c$.
Also, similar solutions appears to the above case when
$\frac{b}{M}=1$ and $|Q|=1.1Q_c$, while one less stable and unstable
exist when $\frac{b}{M}=2$ and $|Q|=1.1Q_c$. However, for
$\frac{b}{M}=5$, same solutions are found either we take $\beta=0.2$
or $\beta=1$ with $|Q|=0.9999Q_c$ and $|Q|=1.1Q_c$. Also a similar
behavior of horizon radius is observed.}
\end{itemize}

\section{Discussion and Conclusions}

In this work, we have formulated BI thin-shell WHs supported with
GCCG and look into their linearize stability analysis via radial
perturbations (preserve the symmetry). We have solved evolution
equation (\ref{14}) of static WHs numerically and used in (\ref{24})
to investigate their stability. We have found static WHs solutions
corresponding to different values of charge, gas exponent
$\beta=0.2,~1$, BI parameter $\frac{b}{M}=1,~2,~5$, $\omega=-5$ and
constants involve in the model. The results are shown in Figures
\textbf{1-6}. The solutions in the light (green and yellow) regions
are represented as (stable and unstable) solutions respectively.
\newpage
\begin{center} {\bf {\small Table 1}} {\small \textbf{Comparison
of BI Static WH Solutions with GCG, MCG and GCCG EoS}}

\vspace{0.2in}
\begin{tabular}{|c|c|c|c|c|c|c|}

\hline{\bf Value of $\beta$}&{\bf EoS}&{$\frac{b}{M}=$}&{\bf
$\frac{|Q|}{M}=0$}
&{\bf $\frac{|Q|}{M}=0.7$}&{\bf $\frac{|Q|}{M}=0.999$}&{\bf $\frac{|Q|}{M}=1.1$}\\
\hline $\beta=0.2$& $GCG$& $1$& $1U$ & $1U$& $2U,1S$& $2U,1S$\\
\hline $\beta=0.2$& $MCG$& $1$& $1U$ & $1U$& $2U,1S$& $2U,1S$\\
\hline $\beta=0.2$& $GCCG$& $1$& $1U$ & $1U$& $2U,1S$& $2U,2S$\\
\hline $\beta=0.2$& $GCG$& $2$& $1U$ & $1U$& $2U,1S$& $2U,1S$\\
\hline $\beta=0.2$& $MCG$& $2$& $1U$ & $1U$& $2U,1S$& $1U$\\
\hline $\beta=0.2$& $GCCG$& $2$& $1U$ & $1U$& $2U,1S$& $2U,2S$\\
\hline $\beta=1$& $GCG$& $1$& $1U$ & $1U$& $1U,1S$& $2U,1S$\\
\hline $\beta=1$& $MCG$& $1$& $1U,1S$ & $1U,1S$& $1U,1S$& $1U$\\
\hline $\beta=1$& $GCCG$& $1$& $1U$ & $1U$& $1U,1S$& $2U,2S$\\
\hline $\beta=1$& $GCG$& $2$& $1U$ & $1U$& $1U,1S$& $1U$\\
\hline $\beta=1$& $MCG$& $2$& $1U,1S$ & $1U,1S$& $1U,1S$& $1U$\\
\hline $\beta=1$& $GCCG$& $2$& $1U$ & $1U$& $1U,1S$& $1U,1S$\\
\hline $\beta=0.2$& $GCG$& $5$& $1U$ & $1U$& $1U$& $1U$\\
\hline $\beta=0.2$& $MCG$& $5$& $1U$ & $1U$& $1U$& $1U$\\
\hline $\beta=0.2$& $GCCG$& $5$& $1U$ & $1U$& $2U$& $1U,1S$\\
\hline $\beta=1$& $GCG$& $5$&$ 1U$ & $1U$& $1U$& $1U$\\
\hline $\beta=1$& $MCG$& $5$& $1U,1S$ & $1U,1S$& $1U$& $1U$\\
\hline $\beta=1$& $GCCG$& $5$& $1U$ & $1U$& $2U$& $1U,1S$\\
\hline
\end{tabular}
\end{center}
In order to see the role played by the GCCG EoS in the existence and
stability of static WH solutions, a comparison of static WH
solutions between GCCG, MCG and GCG is given in the table
\textbf{1}. Eiroa and Aguirre (2012) shows that for small values $b$
and charge, there exists similar results to the
Reissner-Nordstr\"{o}m case (Eiroa, 2009). However, for large value
of $b$, the Einstein-Born-Infeld theory deviating from the
Einstein-Maxwell theoryu, the stable regions are disappear and only
unstable solutions are possible. Sharif and Azam (2014) extended
this work with MCG and found stable as well unstable static WH
solutions even for large value of BI parameter.

In this work, we have constructed the BI thin-shell WHs in the
vicinity of GCCG and their stability. We can see a similar behavior
of static solutions from table \textbf{1} when $|Q|$ is not very
close to $Q_c$ ($0\leq|Q|<{Q_c}$) and for different values of charge
and BI parameter, i.e., only unstable WH solution exists all
considered EoS, except the case of MCG, we have one extra stable
solution for $\beta=1$ corresponding distinct values of $b$.
However, we have found one extra stable static WH solution
corresponding to distinct values of $\beta$ and $b$ with GCCG when
$|Q|\geq{Q_c}$. This fact supports the consistency of results that
extra solutions are present for large value of charge with GCCG
(Sharif and Azam, 2013d). Thus, it is concluded that GCCG and large
value of charge are the most critical factors for the existence of
such extra stable WH solutions.

\end{document}